\documentclass[journal]{IEEEtran}

\usepackage{amsmath,amssymb}
\usepackage{graphicx}
\usepackage{cite}
\usepackage{booktabs}
\usepackage{balance}
\usepackage{stfloats}
\usepackage{needspace}
\usepackage{tikz}
\usetikzlibrary{arrows.meta,positioning}

\begin{document}

\title{Uncertainty-Aware Multi-Task Learning for Joint Modulation Recognition and SINR Estimation}

\author{Kosar~Nourolahi,~Vahid~Ghasemi%

\thanks{K.~Nourolahi is with the Department of Artificial Intelligence,
Kermanshah University of Technology, Kermanshah Province, Iran.
(e-mail: k.norolahi@smail.kut.ac.ir).}%
\thanks{V.~Ghasemi is with the Department of Artificial Intelligence,
Kermanshah University of Technology, Kermanshah Province, Iran.
(e-mail: v.ghasemi@kut.ac.ir).}}

\maketitle

\begin{abstract}
Joint modulation recognition and signal-to-interference-plus-noise
ratio (SINR) estimation can reduce duplicated processing in
intelligent receivers, but the two tasks have different uncertainty
characteristics. This letter proposes an uncertainty-aware multi-task
model that transforms each short normalized in-phase/quadrature
window into 36 deterministic, label-free statistics, learns a shared
representation, and uses task-specific adapters for modulation
classification and heteroscedastic SINR regression. A joint
uncertainty score combines classification entropy and predicted
regression variance to support selective inference. Simulations cover
QPSK, 8PSK, 16QAM, and 64QAM under matched additive white Gaussian
noise/Rayleigh channels and an unseen frequency-selective Rician
channel. Over five independent seeds, the proposed model improves
matched and unseen-channel accuracy over conventional multi-task
learning by 14.86 and 8.61 percentage points, respectively, while
reducing SINR mean absolute error by 1.60 and 1.61 dB.
Confidence-based rejection further lowers modulation error under
channel mismatch.
\end{abstract}


\section{Introduction}
Automatic modulation classification (AMC) and signal-quality
estimation are central inference functions in intelligent radio
receivers. Deep AMC models can learn discriminative representations
from in-phase/quadrature (I/Q) observations and improve robustness to
channel impairments \cite{HuynhThe2021Survey,Kim2021RobustAMC, Norolahi2022Blind}, while
real-time implementations demonstrate the feasibility of
learning-assisted AMC \cite{Kaleem2021RealTime}. Signal-to-noise ratio
estimation has likewise been addressed using learned covariance
representations \cite{Chen2023SNREstimation}. Nevertheless, AMC and
SINR estimation are usually trained as separate tasks even though
both operate on the same received waveform.

Joint modulation and signal-quality determination has previously
relied mainly on engineered descriptors
\cite{Almohamad2018JointSNR,Almohamad2021Dual}. Multi-task learning
has also been applied to related radio-signal tasks, including pulse
detection and modulation classification \cite{Akyon2019MultiTask}.
In our earlier work, modulation fingerprints combined
distributional, spectral, and constellation-geometry features for
AMC \cite{Norolahi2021Fingerprint}. We subsequently used an RBF-SVM
to recognize modulation and derived SINR information from
support-vector samples for link adaptation
\cite{Norolahi2023JointImprovement}. These approaches established
the value of interpretable signal statistics, but they did not
jointly learn task-adaptive representations or quantify prediction
reliability.

Conventional shared-encoder learning can suffer from negative
transfer because modulation recognition is discrete whereas SINR
estimation is continuous, and their difficulty changes differently
with noise, interference, and channel mismatch. Uncertainty-aware AMC
has been studied for unreliable classifications
\cite{Luu2024UncertaintyAMC}, and calibrated prediction has been
considered for wireless learning systems \cite{Cohen2023Calibration}.
However, the cited studies do not jointly couple AMC with continuous
SINR regression or use both output uncertainties for selective joint
inference.

This letter makes three contributions. First, it develops a compact
multi-task architecture that combines a reproducible 36-dimensional
signal representation with a shared encoder and residual task
adapters. Second, it couples modulation cross-entropy with robust
heteroscedastic SINR regression and combines classification entropy
with predicted SINR deviation for selective inference. Third, it
performs a validation-only, five-seed comparison under both matched
channels and a deliberately unseen frequency-selective Rician
channel, including confidence intervals, calibration, classwise
confusion, and coverage-risk behavior.

\section{Problem Formulation}
Let $m\in\{1,\ldots,M\}$ denote the modulation class and
$s_m[\ell]$ the transmitted baseband sequence. The received samples
are modeled as
\begin{equation}
 r[\ell]
 =e^{j(2\pi\nu\ell+\phi)}
 \sum_{p=0}^{P-1}h_p s_m[\ell-p-\tau]
 +i[\ell]+w[\ell],
 \label{eq:rx}
\end{equation}
where $h_p$ is the $p$th channel tap, $\nu$, $\phi$, and $\tau$
denote carrier-frequency, phase, and timing offsets, respectively,
$i[\ell]$ is cochannel interference, and $w[\ell]$ is receiver
noise. The reference SINR is
\begin{equation}
 \gamma
 =10\log_{10}
 \frac{\mathbb{E}\{|\sum_p h_p s_m[\ell-p-\tau]|^2\}}
 {\mathbb{E}\{|i[\ell]+w[\ell]|^2\}}.
 \label{eq:sinr}
\end{equation}
For a window $\mathbf r=[r[0],\ldots,r[L-1]]^{\mathsf T}$, power
normalization gives
$\widetilde{\mathbf r}=\mathbf r/(L^{-1}\|\mathbf r\|_2^2+\epsilon)^{1/2}$.
The supervised set is
$\mathcal D=\{(\widetilde{\mathbf r}_n,y_n,\gamma_n)\}_{n=1}^{N}$,
and the desired mapping is
\begin{equation}
 F_{\Theta}(\widetilde{\mathbf r}_n)
 =\big(\widehat{\mathbf p}_n,
 \widehat{\gamma}_n,\widehat{\sigma}_n^2\big),
 \label{eq:mapping}
\end{equation}
where $\widehat{\mathbf p}_n$ is the modulation probability vector
and $(\widehat{\gamma}_n,\widehat{\sigma}_n^2)$ are the predicted
SINR mean and data-dependent variance. Training samples are generated
from $\Omega_{\rm tr}$, whereas the mismatch set uses channel and
impairment parameters in
$\Omega_{\rm te}\not\subseteq\Omega_{\rm tr}$.

\Needspace{11\baselineskip}
For a confidence threshold $\eta$, let
$\mathcal I_{\eta}=\{n:u_n\leq\eta\}$ contain the retained
predictions. The retained coverage and task risks are
\begin{align}
 C(\eta)
 &=\frac{|\mathcal I_{\eta}|}{N},
 \nonumber\\
 R_{\rm c}(\eta)
 &=\frac{1}{|\mathcal I_{\eta}|}
 \sum_{n\in\mathcal I_{\eta}}
 \mathbf{1}\{\widehat y_n\neq y_n\},
 \nonumber\\
 R_{\rm r}(\eta)
 &=\frac{1}{|\mathcal I_{\eta}|}
 \sum_{n\in\mathcal I_{\eta}}
 |\widehat\gamma_n-\gamma_n|.
 \label{eq:selective_risks}
\end{align}
Thus, selective inference is evaluated by the classification-error
and SINR-MAE risks at comparable retained coverage.

\section{Proposed Uncertainty-Aware Multi-Task Model}
\begin{figure}[t]
\centering
\resizebox{\columnwidth}{!}{%
\begin{tikzpicture}[
node distance=3.8mm,
>=Latex,
box/.style={draw,rounded corners,align=center,minimum height=6.5mm,inner xsep=3mm,font=\scriptsize},
arr/.style={->,thick}]
\node[box] (iq) {Normalized\\I/Q window};
\node[box,right=of iq] (stat) {36 signal\\statistics};
\node[box,right=of stat] (enc) {Shared\\encoder};
\node[box,right=of enc,yshift=4.5mm] (cls) {Class adapter\\$\widehat{\mathbf p}$};
\node[box,right=of enc,yshift=-4.5mm] (reg) {SINR adapter\\$\widehat\gamma,\widehat\sigma$};
\draw[arr] (iq) -- (stat);
\draw[arr] (stat) -- (enc);
\draw[arr] (enc) -- (cls);
\draw[arr] (enc) -- (reg);
\end{tikzpicture}}
\caption{Proposed statistics-based uncertainty-aware multi-task
inference chain.}
\label{fig:architecture}
\end{figure}
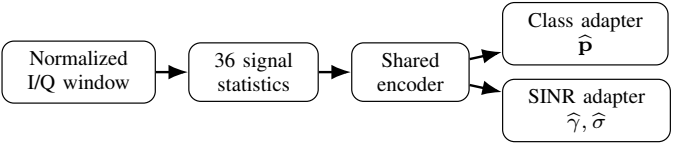

\emph{Statistical representation:}
From each normalized I/Q window, a vector
$\mathbf f\in\mathbb R^{36}$ is computed without labels. The exact
feature allocation is summarized in Table~\ref{tab:features}. Every
feature is standardized using only the training-set mean and
variance, preventing information leakage from validation or test
observations.

\begin{table}[t]
\caption{Deterministic input representation.}
\label{tab:features}
\centering
\scriptsize
\setlength{\tabcolsep}{3.7pt}
\renewcommand{\arraystretch}{0.94}
\begin{tabular}{lc}
\toprule
Feature group & Dimension\\
\midrule
I/Q/amplitude moments (orders 1-4) & 12\\
Complex higher-order moments & 5\\
Phase-increment harmonics (orders 1-4) & 8\\
Lagged correlations ($1,2,4,8$) & 8\\
Spectral entropy/flatness/concentration & 3\\
\midrule
Total & 36\\
\bottomrule
\end{tabular}
\end{table}

\emph{Shared and task-adapted representations:}
As illustrated in Fig.~\ref{fig:architecture}, a layer-normalized
$36\!\rightarrow\!64\!\rightarrow\!48$ encoder produces
$\mathbf z\in\mathbb R^{48}$. To retain shared information while
allowing task specialization, residual adapters generate
\begin{equation}
 \mathbf z_c
 =\frac{\mathbf z+\rho A_c(\mathbf z)}
 {\|\mathbf z+\rho A_c(\mathbf z)\|_2},
 \qquad
 \mathbf z_r
 =\mathbf z+\rho A_r(\mathbf z),
 \label{eq:adapters}
\end{equation}
with $\rho=0.35$. The classifier gives
$\widehat{\mathbf p}=\operatorname{softmax}(W_c\mathbf z_c+\mathbf b_c)$.
The regression head outputs a standardized SINR mean $\widehat\mu$
and log-variance $\widehat v\in[-5,4]$, which are transformed back
to $\widehat\gamma$ and $\widehat\sigma^2$ using the mean and
variance of the training labels.

\emph{Training objective:}
The modulation loss is cross-entropy,
$\mathcal L_{\rm cls}=-N^{-1}\sum_n\log\widehat p_{n,y_n}$.
To retain heteroscedastic uncertainty without allowing a few large
errors to dominate, the SINR term combines Gaussian negative
log-likelihood and smooth-$\ell_1$ regression:
\begin{align}
 \mathcal L_{\rm reg}
 &=\alpha\mathcal L_{\rm nll}
 +(1-\alpha)\mathcal L_{\rm hub},
 \nonumber\\
 \mathcal L_{\rm nll}
 &=\frac{1}{2N}\sum_n
 \left[e^{-\widehat v_n}(t_n-\widehat\mu_n)^2+\widehat v_n\right],
 \nonumber\\
 \mathcal L
 &=\mathcal L_{\rm cls}+\lambda\mathcal L_{\rm reg},
 \label{eq:loss}
\end{align}
where $t_n$ is the standardized SINR target, $\alpha=0.6$, and
$\lambda=0.4$.

\emph{Selective inference:}
The normalized classification entropy
$u_c=-\sum_m\widehat p_m\log(\widehat p_m+\epsilon)/\log M$ is
combined with regression uncertainty as
\begin{equation}
 u
 =0.55u_c
 +0.45\operatorname{clip}(\widehat\sigma/10,0,1).
 \label{eq:uncertainty}
\end{equation}
Predictions are ranked by $u$; rejecting the largest values yields
the coverage-risk tradeoff in \eqref{eq:selective_risks} without
retraining.

\subsection{Complexity and Interpretation}
The statistical front end requires linear-time moment, phase, and
correlation calculations plus one $L$-point fast Fourier transform,
giving $\mathcal O(L\log L)$ complexity and a fixed 36-dimensional
output. Excluding normalization and nonlinear activations, the dense
network requires 16,384 multiply-accumulate operations per
observation. Its 16,798 trainable parameters are 42.7\% fewer than
the 29,302 parameters of the conventional complex-CNN multi-task
baseline. The shared encoder captures common signal-quality and
modulation information, while the residual adapters limit
destructive task interference without duplicating the full encoder.

\section{Simulation Results}
QPSK, 8PSK, 16QAM, and 64QAM are simulated using $L=128$ samples at
nominal SINRs $\{-10,-5,0,5,10,15,20\}$ dB, with each label
uniformly perturbed by $\pm1.25$ dB. Training, validation, and
matched testing use AWGN and flat Rayleigh fading. The unseen set
uses a three-tap frequency-selective Rician channel with factor
$K\in[2,7]$ and wider carrier-frequency, timing, I/Q-imbalance,
direct-current-offset, and interference ranges. The four sets contain
5600, 1680, 4480, and 3360 observations, respectively, with no shared
symbol, channel, noise, or interference realization.

AdamW uses learning rate $1.5\times10^{-3}$, weight decay $10^{-4}$,
batch size 256, gradient-norm clipping at 5, and at most 20 epochs.
Early stopping follows five epochs without improvement. The retained
checkpoint maximizes
$A_{\rm val}-0.02\,\mathrm{MAE}_{\rm val}$ using validation data only.
Neural results are means over seeds
$\{101,202,303,404,505\}$. Baselines are RBF-SVM/SVR on the same
statistics, separate complex CNNs, and a conventional shared-encoder
complex CNN.

\begin{table*}[t]
\caption{Overall performance. Neural results are mean $\pm$ standard deviation over five independent seeds.}
\label{tab:overall}
\centering
\scriptsize
\setlength{\tabcolsep}{4.2pt}
\renewcommand{\arraystretch}{0.94}
\begin{tabular}{llcccc}
\toprule
Test & Method & Accuracy & Macro-F1 & MAE (dB) & RMSE (dB)\\
\midrule
Matched
& RBF-SVM/SVR
& 0.460 & 0.461 & 3.765 & 4.782\\
& Separate CNNs
& $0.335\!\pm\!0.019$ & $0.295\!\pm\!0.015$
& $4.752\!\pm\!0.098$ & $6.024\!\pm\!0.170$\\
& Conventional MTL
& $0.352\!\pm\!0.009$ & $0.297\!\pm\!0.022$
& $4.904\!\pm\!0.303$ & $6.310\!\pm\!0.439$\\
& \textbf{Proposed}
& $\mathbf{0.500\!\pm\!0.004}$ & $\mathbf{0.508\!\pm\!0.004}$
& $\mathbf{3.302\!\pm\!0.062}$ & $\mathbf{4.158\!\pm\!0.045}$\\
\midrule
Unseen Rician
& RBF-SVM/SVR
& 0.355 & 0.357 & 5.153 & 6.564\\
& Separate CNNs
& $0.306\!\pm\!0.011$ & $0.247\!\pm\!0.028$
& $5.968\!\pm\!0.181$ & $7.793\!\pm\!0.288$\\
& Conventional MTL
& $0.321\!\pm\!0.005$ & $0.253\!\pm\!0.016$
& $6.317\!\pm\!0.354$ & $8.218\!\pm\!0.440$\\
& \textbf{Proposed}
& $\mathbf{0.407\!\pm\!0.005}$ & $\mathbf{0.402\!\pm\!0.003}$
& $\mathbf{4.703\!\pm\!0.223}$ & $\mathbf{5.938\!\pm\!0.274}$\\
\bottomrule
\end{tabular}
\end{table*}

Table~\ref{tab:overall} shows that the proposed model improves
matched accuracy over conventional MTL by 14.86 percentage points
and reduces MAE by 1.602 dB. Under unseen Rician fading, the gains
remain 8.61 percentage points and 1.614 dB. The paired 95\%
confidence intervals are $[13.86,15.86]$ and $[8.16,9.07]$
percentage points for matched and unseen accuracy, and
$[1.205,1.999]$ and $[1.018,2.210]$ dB for MAE; all exclude zero.
The corresponding macro-F1 gains are 0.211 and 0.149, while the RMSE
reductions are 2.152 and 2.280 dB. Relative to RBF-SVM/SVR, the
proposed model yields smaller point gains: 4.08/5.16 percentage
points in accuracy and 0.463/0.450 dB in MAE for matched/unseen
testing.

\begin{figure}[t]
\centering
\includegraphics[width=\columnwidth]{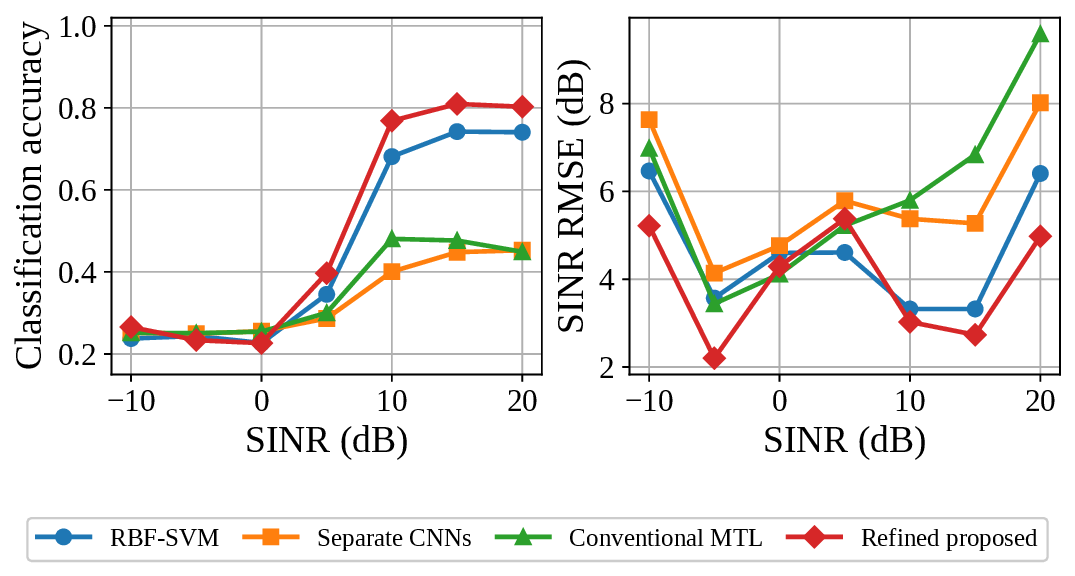}
\caption{Matched-channel performance versus SINR:
(a) modulation-classification accuracy and
(b) SINR-estimation RMSE. Neural curves are means over five
independent seeds.}
\label{fig:joint}
\end{figure}

Fig.~\ref{fig:joint} shows that all classifiers approach the
four-class random level below 0 dB. From 10 dB onward, the proposed
model reaches accuracies of 0.768, 0.810, and 0.803 at 10, 15, and
20 dB, exceeding the RBF-SVM values of 0.681, 0.742, and 0.741. It
also gives the lowest aggregate RMSE, although RBF-SVR is better at
the isolated 5-dB point; hence, no claim of uniform dominance at
every SINR is made.

\begin{figure}[t]
\centering
\includegraphics[width=\columnwidth]{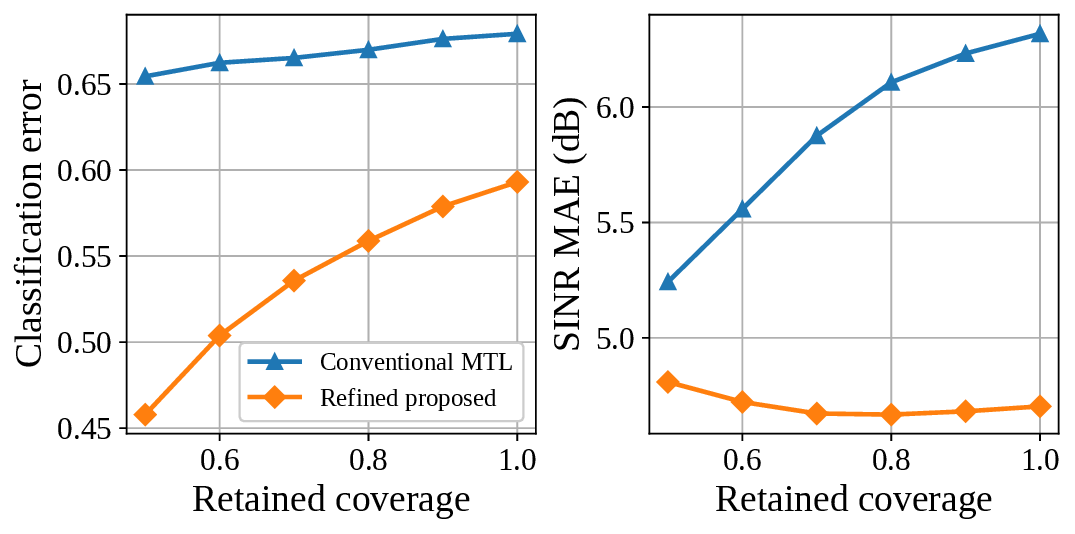}
\caption{Selective inference under unseen Rician fading:
(a) modulation-classification error and
(b) SINR-estimation MAE versus retained coverage.}
\label{fig:selective}
\end{figure}

Fig.~\ref{fig:selective} evaluates the uncertainty score under
mismatch. At full coverage, the proposed and conventional MTL
classification errors are 0.593 and 0.679. Retaining only the 50\%
most confident observations reduces them to 0.458 and 0.654,
respectively. The proposed SINR MAE remains lower at every coverage,
including 4.809 versus 5.242 dB at 50\%. Its minimum is approximately
4.668 dB near 80\% coverage, showing that uncertainty-based rejection
is substantially more effective for AMC than for SINR regression.

\begin{figure}[!t]
\centering
\includegraphics[width=0.98\columnwidth]{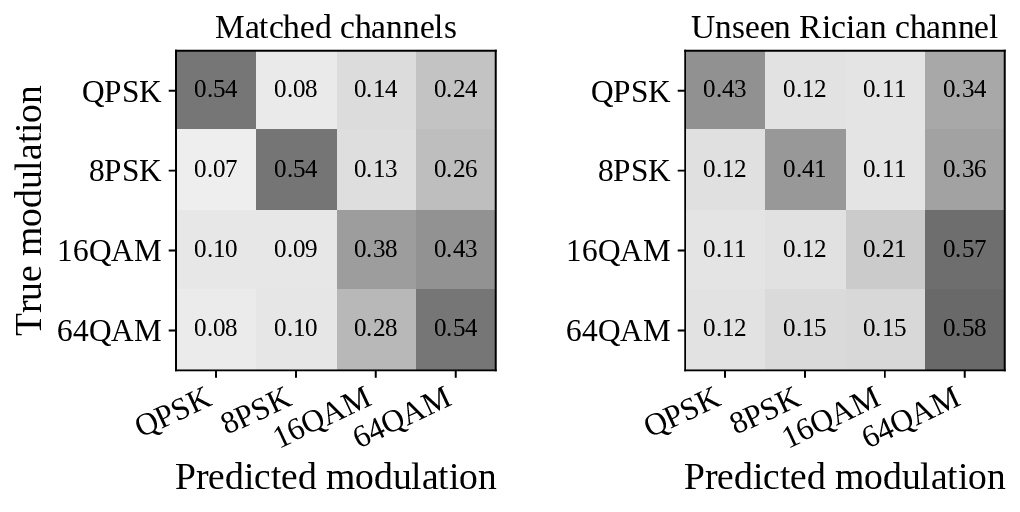}
\caption{Row-normalized confusion matrices of the proposed method
under (a) matched channels and (b) the unseen frequency-selective
Rician channel.}
\label{fig:confusion}
\end{figure}

Fig.~\ref{fig:confusion} clarifies the class-dependent behavior.
Under matched testing, QPSK, 8PSK, and 64QAM each obtain a diagonal
probability of 0.54, whereas 16QAM reaches 0.38 and is most often
confused with 64QAM (0.43). Under the unseen Rician channel, the
diagonal probabilities become 0.43, 0.41, 0.21, and 0.58,
respectively. The dominant mismatch error is the assignment of
16QAM to 64QAM (0.57), while 64QAM remains the most robust class.
This behavior indicates that the principal unresolved ambiguity is
between the two amplitude-phase constellations rather than between
PSK and QAM families.

\subsection{Architecture Selection, Calibration, and Limitations}
A validation-only comparison also examined a raw-I/Q and statistics
hybrid. As shown in Table~\ref{tab:selection}, the final
statistics-based model is slightly more accurate and consistently
reduces SINR MAE. Validation also selected zero weight for the tested
prototype-alignment term; therefore, neither the hybrid branch nor an
unsupported prototype regularizer is retained in the final method.

\begin{table}[t]
\caption{Validation-based architecture selection. Entries are
accuracy/MAE (dB).}
\label{tab:selection}
\centering
\scriptsize
\setlength{\tabcolsep}{3.5pt}
\renewcommand{\arraystretch}{0.94}
\begin{tabular}{lcc}
\toprule
Model & Matched & Unseen Rician\\
\midrule
Raw-I/Q hybrid & 0.494/3.486 & 0.403/4.934\\
Final proposed & \textbf{0.500/3.302} & \textbf{0.407/4.703}\\
\bottomrule
\end{tabular}
\end{table}

The proposed classifier has expected calibration errors of 4.52\%
and 7.59\% for matched and unseen testing, respectively. Its nominal
90\% SINR interval covers 90.8\% of matched observations but only
76.4\% under Rician mismatch. Thus, the joint uncertainty score
remains useful for ranking observations, while absolute interval
calibration degrades under distribution shift. The strongest gains
occur in the moderate- and high-SINR regions, where the deterministic
statistics expose constellation geometry and higher-order structure
with fewer samples than the raw-I/Q baselines. At very low SINR, all
classifiers remain close to random guessing, indicating that the
method does not create separability when the observation contains
insufficient information. The conclusions are limited to simulated
impairments, four modulation formats, and $L=128$; measured I/Q data
and broader open-set conditions remain necessary for external
validation.

\Needspace{16\baselineskip}
\section{Conclusion}
A statistics-based uncertainty-aware multi-task model was developed
for joint AMC and SINR estimation. Shared learning with residual task
adapters and robust heteroscedastic regression improved both tasks
under matched and unseen channels, while entropy-variance ranking
enabled selective inference. The results also expose two limitations:
16QAM-to-64QAM confusion remains substantial under mismatch, and
prediction-interval calibration degrades outside the training
distribution. Future work should address these limitations using
measured I/Q data and distribution-aware recalibration.

\bibliographystyle{IEEEtran}
\bibliography{references_ieee_letter}

\end{document}